\documentclass{article}
\usepackage[T1]{fontenc} % add special characters (e.g., umlaute)
\usepackage[utf8]{inputenc} % set utf-8 as default input encoding
\usepackage{ismir,amsmath,cite,url}
\usepackage{graphicx}
\usepackage{color}

\def\ourtool{MiRA} % Name of our tool
\usepackage{xcolor}
\newcommand{\roser}[1]{\textcolor{black}{#1}}
\usepackage{url}
\usepackage{graphicx}
\usepackage{subcaption}
\usepackage{rotating}
\usepackage{tabularray}
\usepackage{array}
\usepackage{booktabs}
\usepackage{amssymb}

\usepackage{lineno}
\title{Towards Assessing Data Replication in Music Generation with Music Similarity Metrics on Raw Audio}
\multauthor
{Roser Batlle-Roca$^1$ \hspace{1cm} Wei-Hsiang Liao$^2$ \hspace{1cm} Xavier Serra$^1$} { \bfseries{Yuki Mitsufuji$^3$ \hspace{1cm} Emilia Gómez$^{1,4}$}\\
 $^1$ Music Technology Group, Universitat Pompeu Fabra, Spain \hspace{0.25cm}
$^2$ Sony AI, Japan \hspace{0.25cm}$^3$ Sony AI, USA\\
$^4$ Joint Research Centre, European Commission, Spain\\
{\tt\small roser.batlle@upf.edu\vspace{-0.3cm}}
}

\begin{document}

\maketitle

\begin{abstract}
Recent advancements in music generation are raising multiple concerns about the implications of AI in creative music processes, current business models and impacts related to intellectual property management. A relevant discussion and related technical challenge is the potential replication and plagiarism of the training set in AI-generated music, which could lead to misuse of data and intellectual property rights violations. 
To tackle this issue, we present the Music Replication Assessment (MiRA) tool: a model-independent open evaluation method based on diverse audio music similarity metrics to assess data replication. We evaluate the ability of five metrics to identify exact replication by conducting a controlled replication experiment in different music genres using synthetic samples. Our results show that the proposed methodology can estimate exact data replication with a proportion higher than $10\%$. 
By introducing the MiRA tool, we intend to encourage the open evaluation of music-generative models by researchers, developers, and users concerning data replication, highlighting the importance of the ethical, social, legal, and economic consequences. Code and examples are available for reproducibility purposes.\footnote{\url{https://github.com/roserbatlleroca/mira}}
\end{abstract}
\vspace{-0.15cm}
\section{Introduction}\label{sec:introduction}
%%%%%%%%%%%%%%%%%%%%%%
% Context & problem
%%%%%%%%%%%%%%%%%%%%%%
Significant advancements in generative algorithms for digital art creation are challenging the role of artificial intelligence (AI) in artistic practices. 
Regarding generative AI in the music domain, there is an increasing discussion related to the use of computational tools in music creative processes \cite{Carnovalini2020}, the effects on artists' work, existing listening experiences and business models, and the impacts on intellectual property (IP) management \cite{Gomez2018, Strum2019}. 
A key point is the potential replication 
and plagiarism of the training set in AI-generated music \cite{Strum2019, Yin2022}, which can lead to data misuse and IP violations. 

%The inherent opaque nature of music generation models challenges tracing 
%references in the training set used in 
%AI-generated music, limiting interpretation of whether generated samples contain replicated fragments. 
The inherent opaque nature of music generation models complicates tracing replications or references to training set samples in AI-generated music, limiting the interpretation of whether generated samples contain replicated fragments.
In addition, diffusion models, one of the most popular generative AI architectures, %have been shown a tendency to memorise %
tend to memorise 
and replicate training data \cite{Carlini2023a, Carlini2023b, bralios2023}. 
Understanding the behaviour of these models has become critical to address legal issues \cite{Wang2023}, especially when dealing with data protected by IP rights.
This is significant in the music domain as the vast majority of music is protected by authorship and copyright.

%%%%%%%%%%%%%%%%%%%%%%
% Research gap & Goal
%%%%%%%%%%%%%%%%%%%%%%
Despite multiple claims emphasising the importance of assessing music-generative algorithms, there is a lack of evaluation tools directly focused on detecting data replication 
based on raw audio. 
Considering this research gap, the present investigation is motivated by two main questions: 
\vspace{-0.1cm}
\begin{itemize}
    \item Are audio-based music similarity metrics suitable to assess
    data replication in AI-generated music? 
    \vspace{-0.1cm}
    \item Can we propose an open model-agnostic evaluation method and tool found on diverse audio-based music similarity metrics?
\end{itemize}
\vspace{-0.1cm}

Thus, this work proposes \roser{assessing the effectiveness of} five music similarity metrics\footnote{Hereafter, music similarity metrics refer to audio-based metrics.} (four state-of-the-art widely-used and a novel one) 
in estimating exact data replication in music.
%
%%%%%%%%%%%%%%%%%%%%%%
% Scope of the study
%%%%%%%%%%%%%%%%%%%%%%
%
We review the 
implications of potential data replication in AI-generated music 
(Section \ref{sec:background}) \roser{and} %
present our experimental setup, including the selected music similarity metrics and specific methodology to control and estimate \roser{exact }data replication (Section \ref{sec:experiment}). 
We analyse metrics' behaviour in different music materials (Section \ref{sec:res_metric}), aiming to assess later their data replication detection sensitivity (Section \ref{sec:res_sensitivity}). The proposed methodology is implemented in
tool 
MiRA (Music Replication Assessment), which computes music similarity between reference
and target samples to obtain global and per-pair distances
(Section \ref{sec:tool}). Finally, we discuss \roser{our research's} insights, limitations and future perspectives (Section \ref{sec:conclusions}). 

%%%%%%%%%%%%%%%%%%%%%%
% Contribution
%%%%%%%%%%%%%%%%%%%%%%
By introducing
\ourtool~tool, we advance towards the assessment of 
data replication in AI-generated music using
similarity metrics,
contributing to open evaluation methods for accessibility for 
researchers, developers and users.  
We strive to raise awareness, 
detect and prevent misappropriation of training sets, and hope to motivate research on these issues. 

\section{Background and Related Work}
\label{sec:background}
\subsection{Implications of potential data replication in AI-generated music}
Music-generative AI is advancing rapidly with novel high-quality models driven by a strong push from the industry, which is encouraging a suitable environment for real-world deployment. Yet, music generation algorithms bring significant concerns regarding their ethical, social, legal and economic implications. A key challenge is the potential data replication in AI-generated music---inquiring whether a generative model extracts and copies fragments from the training data and whether AI-generated music can be considered novel and original \cite{Strum2019, Yin2022}. This issue is further complicated by the implications derived concerning data misuse and IP violations such as copyright infringement. 
\roser{Moreover, diffusion models, one of the most popular architectures for generative AI, present high risks of data replication as they have shown a tendency to memorise their training data \cite{Carlini2023a, Carlini2023b, bralios2023}.}
In the image generation domain, Somepalli et al. \cite{Somepalli2022} demonstrate instances where \roser{generated images with }diffusion models \roser{contain object-level copies of} their training data. Based on image retrieval frameworks, they compare generated images with training samples and detect when content has been replicated. \roser{Similarly, Carlini et al. \cite{Carlini2023a} demonstrate that diffusion models memorize and reproduce images from their training data. }

Memorising training data and potential IP violations is highly \roser{under-discussed }
in music generative models literature, despite being one of generative AI's main negative ethical implications in the 
music domain 
\cite{Barnett2023}.  
However, the recently proposed music generative model \textit{MusicLM} \cite{agostinelli2023} has been refrained from releasing due to the ethical risks and potential work replication. In addition, \textit{MusicLDM} \cite{Chen2023} acknowledges potential issues linked to data replication and plagiarism and, to address them, proposes two beat-synchronous mix-up strategies for data augmentation. The exemplified initiatives underscore the relevance of considering and addressing the ethical implications of these algorithms.  

\subsection{Evaluation methodologies in music generation}
%%%%%%%%%%%%%%%%%%%%%%
% SOTA Evaluation
% methods for MusGenAI
%%%%%%%%%%%%%%%%%%%%%%
Xiong et al. \cite{Xiong2023} present a survey on music generation evaluation methodologies divided into objective, subjective and combined approaches. They highlight a current claim in finding a standardised proper method that aligns with all stakeholders, from developers to musicians and music listeners. However, even if multiple evaluation methodologies exist for music generation models, the literature highlights a lack of evaluation methodologies focused on assessing data replication and the originality of AI-generated music \cite{Yin2022, batlle2023}. \roser{In the symbolic domain, }Yin et al. \cite{Yin2022} introduce the \textit{originality score} to measure the extent to which an algorithm might be copying from the training set.
Nonetheless, there is a growing interest in models outputting directly 
audio music instead of symbolic representations. Thus, a research gap exists in detecting data replication in AI-generated music based on raw audio. 

A recent work by Barnett et al. \cite{Barnett2024} proposes a framework based on two music audio embeddings to assess the similarity between the training data and AI-generated samples \roser{for understanding }training data attribution. Their approach, %is 
based on VampNet \cite{flores2023}, 
\roser{computes }cosine distance on embeddings obtained from CLMR (Contrastive Learning of Musical Representations) \cite{Spijkervet2021} and CLAP (Contrastive Language-Audio Pretraining) \cite{wu2023}. 

Our perspective is that \roser{combining} 
metrics based on audio embeddings, 
acoustic qualities, and features capturing music characteristics, such as chord progression or tonal similarity,  
provides a comprehensive assessment of potential data replication in AI-generated music. 
In this study, we aim to validate the effectiveness of five music similarity metrics and build an open tool to assess exact data replication in AI-generated music using these metrics.

\vspace{-0.2cm}
\section{Forced-Replication Experiment}
\label{sec:experiment}
\subsection{Audio Music Similarity Metrics}
For this study, we consider five music similarly metrics: four state-of-the-art approaches and a novel one, 
%four standard and a novel one,
covering a diversity of characteristics. %, from audio embeddings to state-of-the-art 
%metrics. 
We here describe the metrics (summarised in Table \ref{tab:summarymetrics}) and methods used to implement them.\footnote{Two of the metrics rely on Essentia implementation. Essentia is an open-source library and tools for audio and music analysis, description and synthesis, developed in the Music Technology Group at Universitat Pompeu Fabra: \url{https://essentia.upf.edu}.}

\textbf{Cover Song Identification (CoverID)} 
\cite{Serra2009, Serra2008, Serr2008TransposingCR}: Cover song identification is a task aiming to detect whether two music recordings are based on the same composition, accounting for variations in
tempo, structure, and instrumentation while keeping a similar melodic or harmonic line. CoverID relies on pitch-content features and local alignment. To obtain CoverID distance, we use the implementation available in Essentia.\footnote{\url{https://essentia.upf.edu/reference/std_CoverSongSimilarity.html}}
A low CoverID value suggests substantial composition similarity between the two analysed music samples. 

\textbf{Kullback-Leibler (KL) divergence}:
This metric provides a non-symmetric statistical measurement between reference and target probability distributions relative to their entropy. 
KL divergence has been 
employed in the literature to estimate similarity in music (e.g. \cite{Hoffman2008, Schnitzer2010}), and more recently, 
to assess automatic music generation prompt adherence (e.g. \cite{evans2024}). We aim to  
explore its capabilities to estimate data replication in music samples. 
To obtain probability distributions, 
\roser{we use the PaSST audio classifier proposed in Koutini et al. \cite{Koutini2022}, trained on Audioset.}
This methodology aligns with common practice in the literature, such as in \textit{AudioGen} \cite{kreuk2023} and \textit{MusicGen} \cite{copet2023} to obtain the probabilities of the labels in their audio and music samples. 
To avoid the non-symmetry of KL divergence, we compute reference to target and target to reference KL divergence and, subsequently, average both results to obtain symmetric KL divergence. Low KL divergence indicates a closer similarity between distributions.

\begin{table}
\centering
\caption{Summary of the considered music similarity metrics, indicating whether values correspond to higher or lower similarity ($\downarrow$/$\uparrow$).}
\label{tab:summarymetrics}
\resizebox{\linewidth}{!}{%
\begin{tabular}{>{\centering\hspace{0pt}}m{0.356\linewidth}|>{\hspace{0pt}}m{0.563\linewidth}} 
\toprule
\textbf{Metric}                    & \multicolumn{1}{>{\centering\arraybackslash\hspace{0pt}}m{0.563\linewidth}}{\textbf{Description}}       \\ 
\hline
\textbf{CoverID} ($\downarrow$)          & \begin{tabular}[c]{@{}l@{}}%Estimates 
Musical composition similarity\\based on music-specific\\
characteristics.\\\end{tabular}  \\ 
\hline
\textbf{KL divergence} ($\downarrow$)    & \begin{tabular}[c]{@{}l@{}}%Quantifies 
Differences in distributions from\\ an audio classifier. \end{tabular}  \\ 
\hline
\textbf{CLAP} ($\uparrow$)       & \begin{tabular}[c]{@{}l@{}}Distance between embeddings\\from a music pre-trained model. \end{tabular}  \\ 
\hline
\textbf{DEfNet} ($\uparrow$)     & \begin{tabular}[c]{@{}l@{}}Novel metric based on distance\\ between embeddings from a \\ contrastive learning model for \\music similarity. ~\end{tabular}  \\ 
\hline
\textbf{FAD} ($\downarrow$)  & \begin{tabular}[c]{@{}l@{}}Distance between embeddings \\based on CLAP music model.~\end{tabular}  \\
\bottomrule
\end{tabular}
}
\vspace{-0.4cm}
\end{table}

\textbf{Contrastive Language-Audio Pretraining (CLAP) score} 
\cite{wu2023}: CLAP embeddings\footnote{\url{https://github.com/LAION-AI/CLAP}} allow \roser{to obtain} latent representations of 
audio or text by conditioning information. For instance, \textit{MusicLDM} \cite{Chen2023} uses this metric to assess the novelty in text-to-music generations. 
To compute the CLAP score between two music samples, we extract the audio embeddings \roser{from the} %based on the 
pre-trained music model\footnote{Checkpoints: \texttt{music\_audioset\_epoch\_15\_esc\_90.14.pt}.} for each one and compute the cosine distance between them. A high CLAP score indicates a high similarity between the two music samples. 

\textbf{Discogs-EffNet (DEfNet) score}: %\cite{alonso2020}: 
In addition to state-of-the-art distances between audio embeddings, we incorporate a novel approach based on Essentia models \cite{alonso2020}. Essentia's Discogs-EffNet model\footnote{\url{https://essentia.upf.edu/models.html#discogs-effnet}} provides music audio embeddings trained on Discogs 
metadata with contrastive learning purposes for music similarity. We consider DEfNet score to
observe \roser{the effectiveness of embeddings} of a model trained for a music-related task on 
estimating data replication.  
\roser{Embeddings are extracted} based on track self-supervised annotations\footnote{Embeddings extracted with weights \texttt{discogs\_track\_} \texttt{embeddings-effnet-bs64-1.pb}.} and compute the cosine distance between reference and target samples. 
A high DEfNet score reveals high track similarity. 

\textbf{Fréchet Audio Distance (FAD)} \cite{Kilgour2019, Gui2023}: FAD is an \roser{adaptation} of Fréchet Inception Distance (FID) \roser{for music, comparing }%in the image domain. It compares 
embedding distributions of a reference and a target set, based on the ViGGish model \cite{hershey2017}.
Nonetheless, a recent study by Gui et al. \cite{Gui2023} 
questions whether VGGish is the optimal model for FAD computation for music generation evaluation.
They propose a tool kit\footnote{\url{https://github.com/microsoft/fadtk}} with multiple models to obtain more accurate embeddings to assess AI-generated music when calculating FAD. Consequently, we implement the adapted version of FAD using the CLAP audio music pre-trained model. 
A low FAD score indicates a high resemblance between the compared music samples.

\vspace{-0.3cm}
\subsection{Experimental Approach}
To validate the effectiveness 
of the selected  
 music similarity metrics \roser{in} 
 detecting exact data replication, we carried out a controlled forced-replication experiment with synthetic data, i.e. \roser{replicating }
 music excerpts into another song \roser{under controlled conditions}. 
 \roser{Synthetic data guaranteed }
 that the analysed music samples contained 
 copied instances, limiting our scope to 
 exact data replication. 
 
 \roser{For this experiment, we use an in-house dataset of 30-second audio previews from the Spotify API\footnote{\url{https://developer.spotify.com/documentation/web-api}}, composed of over 18,000 samples and 24 music genre classes.
We \roser{focus on}  
six music genre classes defined by Spotify API internal class labels: \textit{heavy metal}, \textit{afrobeats}, \textit{techno}, \textit{dub}, \textit{cumbia} and \textit{bolero}. \roser{These genres were chosen for their }diverse 
musical compositions and elements, allowing us to examine the metrics across multiple scenarios. This selection was supported using ChatGPT, which affirmed that these genres have distinct musical characteristics.
%his selection aligns with ChatGPT's perspective on genres with distinct musical characteristics.
}

\begin{figure}[h!]
    \centering
    \includegraphics[width=.51\textwidth]{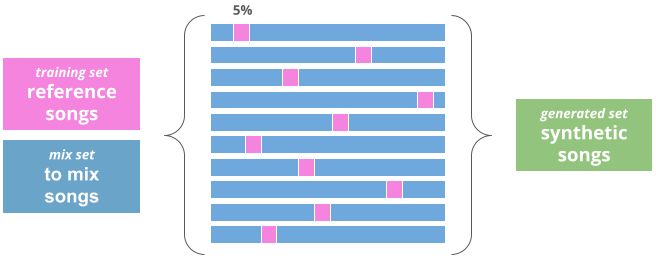}
    \caption{Synthetic data procedure with $5\%$ of replication. }
    \label{fig:syntheticdata}
\end{figure}

We divide data into three groups: (1) \textbf{reference set}: acting as training data, (2) \textbf{target set}: composed of synthetic data, representing AI-generated music, and (3) \textbf{mixture set}: containing different songs from the reference set but from the same music genre to build synthetic data. 
 Synthetic data with replication contains a controlled percentage of copy from a song in our reference set: \roser{$5\%$ (1.5s), $10\%$ (3s), $15\%$ (4.5s), $25\%$ (7.5s) and $50\%$ (15s)}. A synthetic sample is created by introducing the copied proportion 
 at a random point of a music sample in the mixture set.  %of the experiment. 
 We create 10 samples with a proportion of replication per song in the reference set. Figure \ref{fig:syntheticdata} illustrates the procedure to build synthetic data with $5\%$ of replication. 
For each music genre, the reference and mixture sets are composed of 400 songs each. Thus, the target set comprises 4,000 \roser{(400 x 10)} songs per percentage of replication for each genre. Music samples are 30 seconds long as currently it is the common length in full song composition music generative models. 

We assess each metric
for all the songs within the reference set \roser{against themselves to establish a baseline} (\roser{400 x 400 =} 160,000 per-pair evaluations). Then, we compute them for each reference song and its copied instances \roser{to only consider } 
cases with exact data replication (4,000 per-pair evaluations).
Our experiment considers 120,000 samples of synthetic data (approximately 167h of music with a proportion of data replication).

\section{Results}

\subsection{Analysing metric behaviour} 
\label{sec:res_metric}
\begin{figure}[h!]
    \centering
    \includegraphics[width=0.48\textwidth]{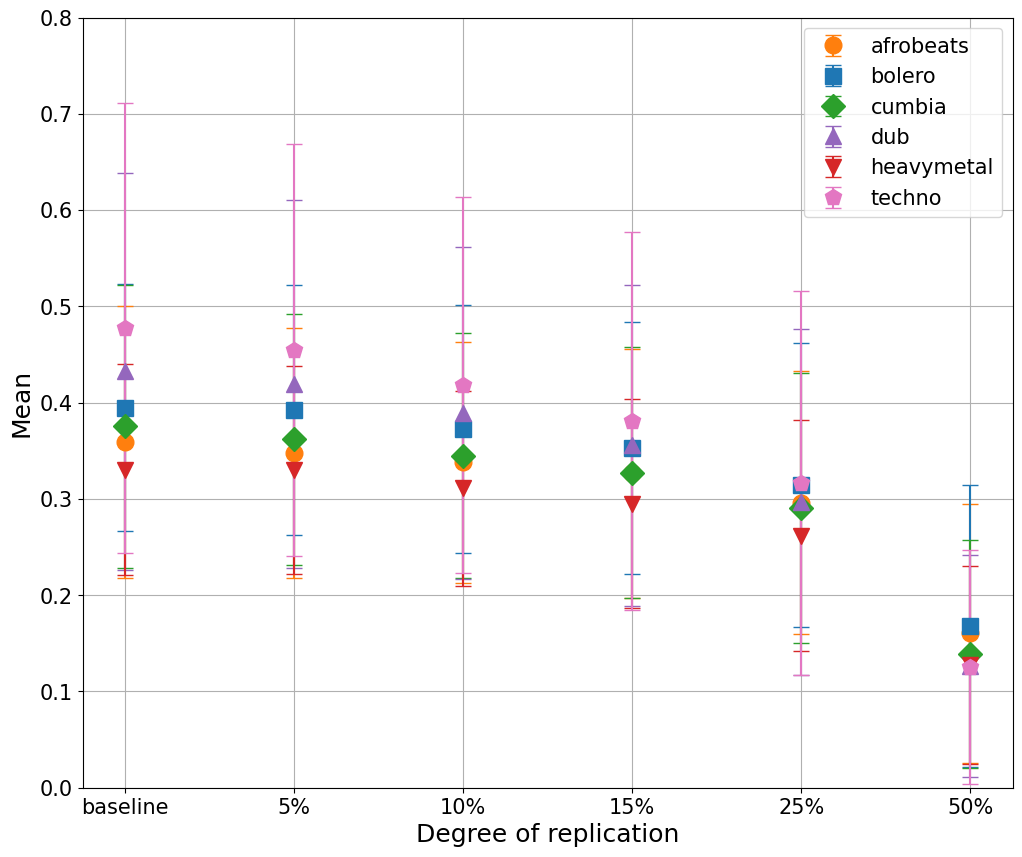}
    \caption{CoverID ($\downarrow$) }
    \label{fig:coverid}
\end{figure}

\begin{figure}[h!]
    \centering
    \includegraphics[width=0.48\textwidth]{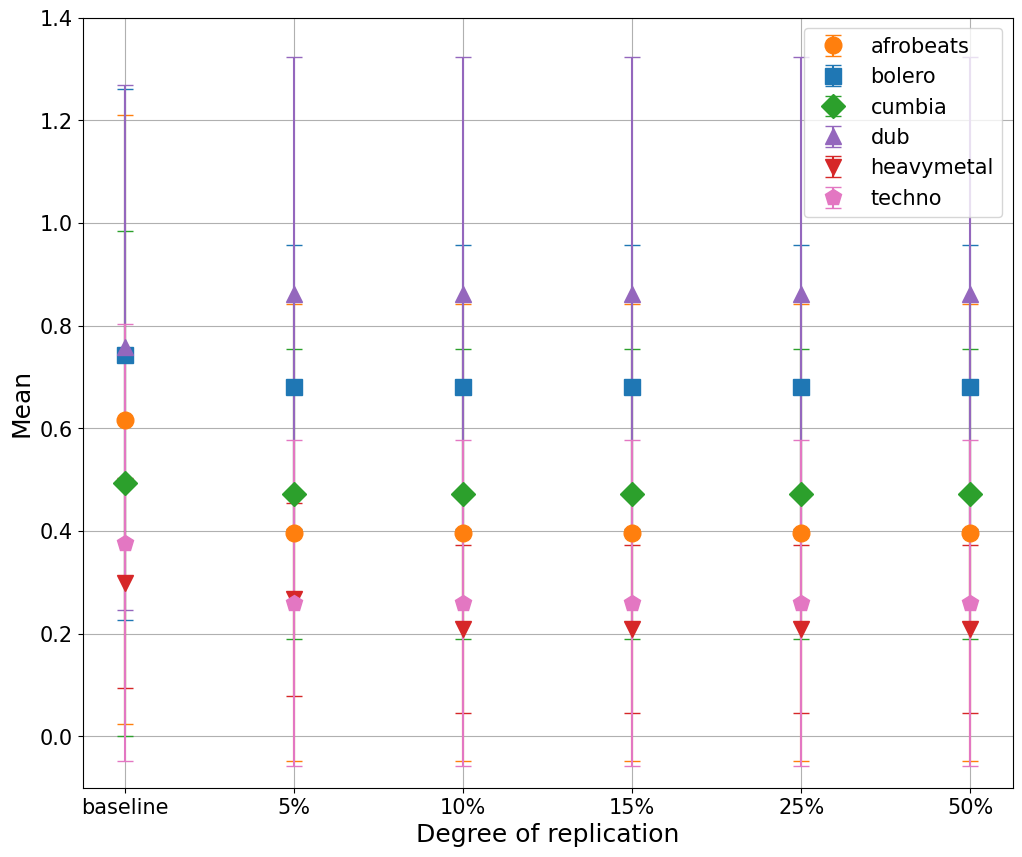}
    \caption{KL divergence ($\downarrow$) }
    \label{fig:KL}
\end{figure}

\begin{figure}[h!]
    \centering
    \includegraphics[width=0.48\textwidth]{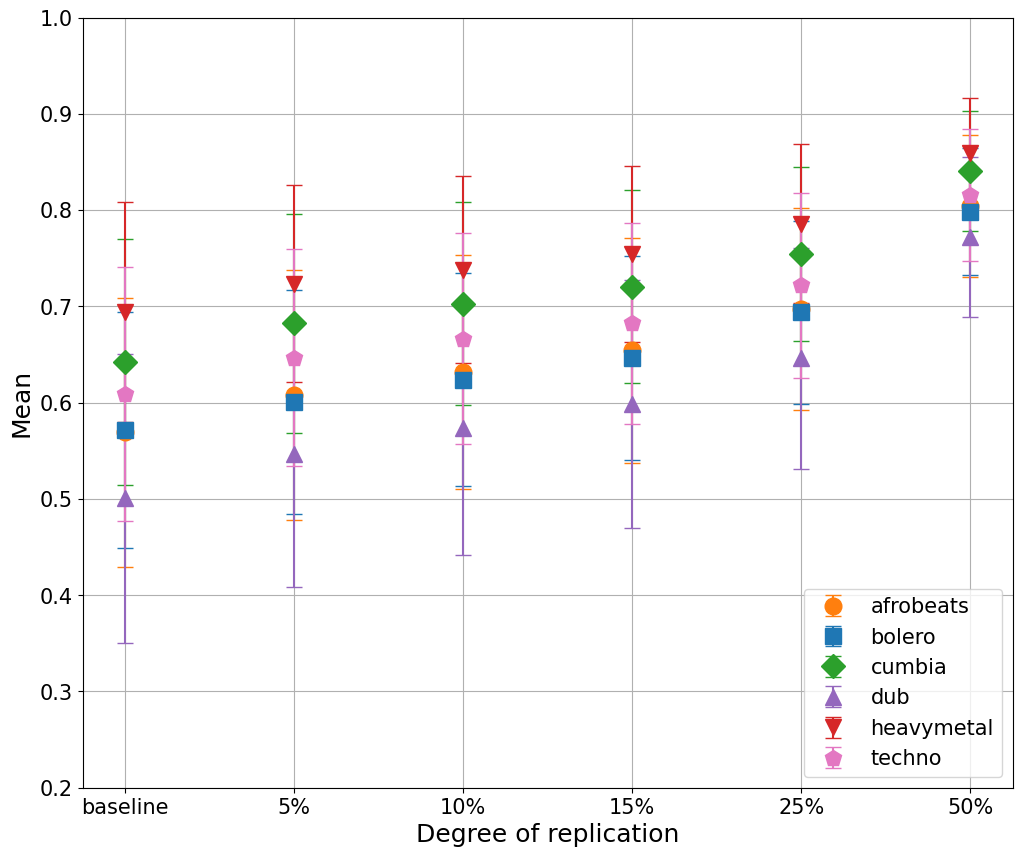}
    \caption{CLAP ($\uparrow$) }
    \label{fig:CLAP}
\end{figure}

\begin{figure}[h!]
    \centering
    \includegraphics[width=0.48\textwidth]{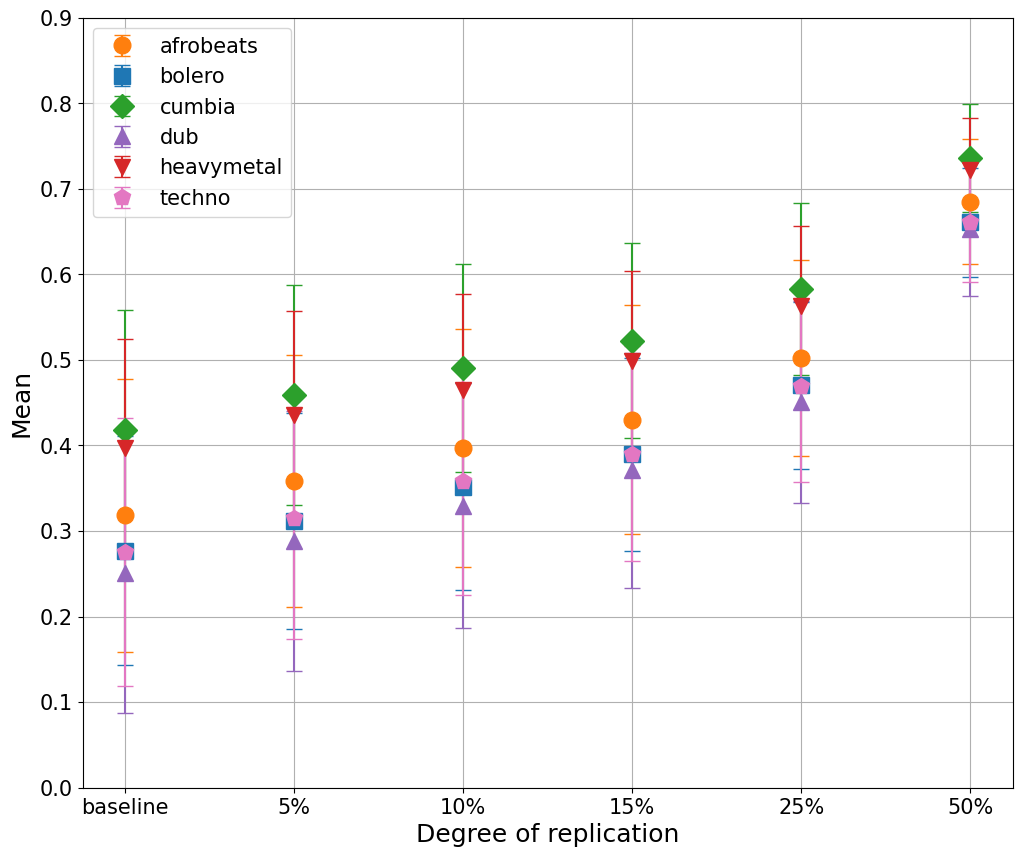}
    \caption{DEfNet ($\uparrow$) }
    \label{fig:DEFNet}
\end{figure}

\begin{figure}[h!]
    \centering
    \includegraphics[width=1\linewidth]{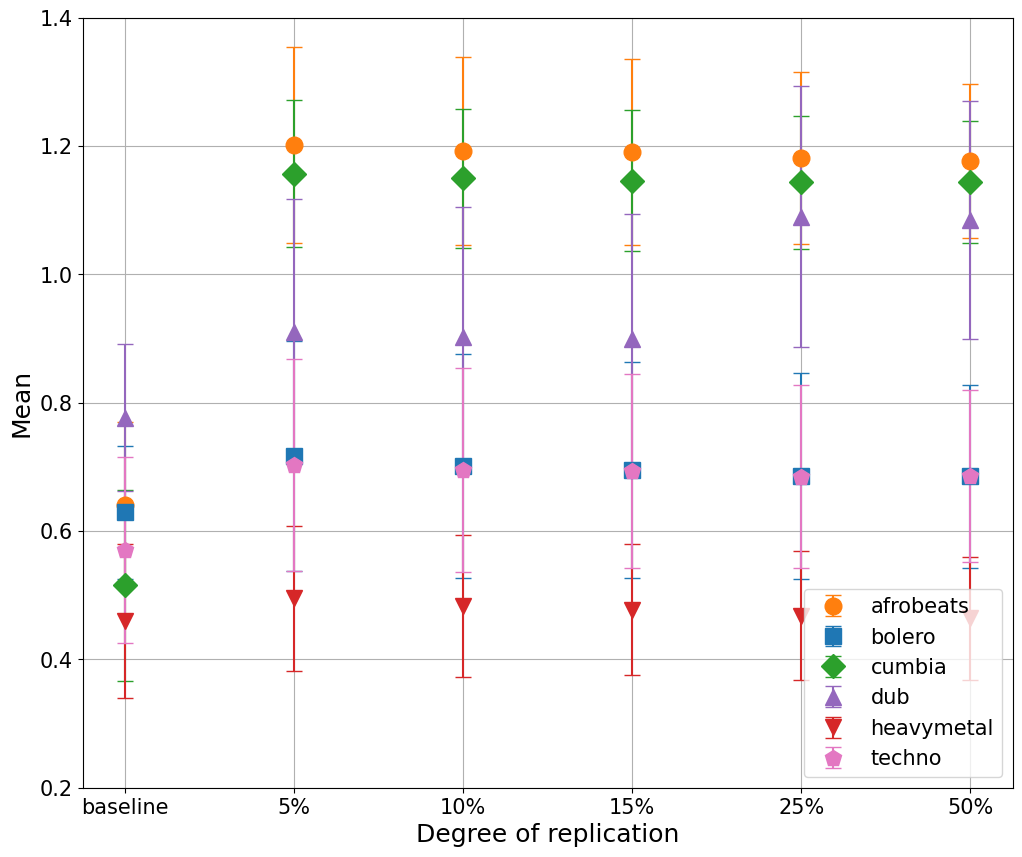}
    \caption{FAD ($\downarrow$) }
    \label{fig:FAD}
\end{figure}

Figures \ref{fig:coverid}, \ref{fig:KL}, \ref{fig:CLAP}, \ref{fig:DEFNet} and \ref{fig:FAD} depict the average $\mu$ and standard deviations $\sigma$ of the different metrics per degree of replication and music genre.
We observe a steady and similar behaviour by three metrics (CoverID, CLAP and DEfNet) through all studied music genres, showing higher similarity values for cases with higher replication levels (i.e., $50\%$). 
Standard deviation decreases with increasing replication level, which suggests less disparity within the analysed pairs. These three metrics show the sensitivity\footnote{\roser{\textit{Sensitivity} is understood} as the capability to differentiate between degrees of replication.}~to estimate data replication. 

Instead, KL divergence presents a different behaviour with very similar values of $\mu$ and $\sigma$ for all degrees of replication. Some sensitivity is observed in all music genres, except for \textit{dub}, where the baseline mean $\mu_b$ is smaller than in replication cases $\mu_r$, despite the standard deviation being higher ($\mu_{b}$=0.757, $\sigma_{b}$=0.511; $\mu_{r}$=0.862, $\sigma_{r}$=0.462). Thus, 
KL divergence demonstrates the capability of detecting replication but is ineffective in distinguishing between degrees of replication. 

Contrasting with \roser{the other} metrics, FAD based on CLAP music embeddings completely differs from them. On the one side, its behaviour is inconsistent as it exhibits fluctuating trends for the different examined cases. 
On the other side, it fails to detect data replication. A higher similarity value (low FAD) is always obtained for the baseline. Instead, for the different degrees of replication, higher FAD is achieved. Consequently, FAD based on CLAP music embeddings does not appear to be a suitable metric to assess exact data replication in music samples. 

By analysing the metrics' behaviour, we could directly conclude that CoverID, KL divergence, CLAP and DEfNet are suitable for our posed research aim. However, further exploration is required before determining their ability to detect replication and 
degree of replication. We delve into this analysis in the next subsection. 

\subsection{Assessing data replication detection sensitivity}
\label{sec:res_sensitivity}
In this section, we complement the previous analysis 
with an assessment of statistical differences. Because our data is not normally distributed and variance is heterogeneous, \roser{the Kruskal-Wallis test \cite{KruskalWallis1952} is the} 
most adequate statistical analysis to examine our results, \roser{as} is non-parametric, does not rely on normality and handles unequal sample sizes. We perform the Kruskal-Wallis test on CoverID, KL divergence, CLAP and DEfNet.
Significant statistical differences ($p < 0.05$) are observed across all music genres and degrees of replication, consistent with our earlier findings.

Nonetheless, the insight of this analysis relies on the pairwise comparisons between the baseline and different degrees of replication. 
CoverID pairwise comparison reveals a statistically significant difference between the baseline and the $5\%$ replication degree for \textit{afrobeat}, \textit{cumbia} and \textit{techno}. For the three other music genres, this happens for a $10\%$ replication degree. Then, statistical significance also appears in pairwise comparisons of different degrees of replication. 
We can derive that CoverID is sensible for $10\%$ of replication, and in some cases at $5\%$. 
When considering KL divergence, pairwise comparison depicts a statistically significant difference between the baseline and the $5\%$ replication degree. Between degrees of replication, no statistical significance is revealed for any pairwise comparison, except for \textit{heavy metal} between $5\%$ and the other replication degrees. % ($p=1.00$). 
Regarding the CLAP and DEfNet, a significant difference already appears when comparing the baseline against the samples with $5\%$ replication, indicating that these metrics are sensitive to 1.5 seconds of replication. In all cases, a notable difference emerges among the levels of replication, enhancing the sensitivity of these metrics' detection capabilities. They demonstrate sensitivity to varying replication degrees. 

Withal, this statistical analysis sustains the validity of these four metrics to assess exact data replication in the training set and determines their degree of sensitivity. 

\section{Music Replication Assessment tool}

\label{sec:tool}
%%%%%%%%%%%%%%%%%%%%%%
% Introduce tool &
% describe structure 
%%%%%%%%%%%%%%%%%%%%%%

Derived from the presented experiment, we implement the proposed methodology into an evaluation tool. We introduce the Music Replication Assessment (\ourtool) tool: an open evaluation method based on four diverse raw audio music similarity metrics. 

MiRA computes music similarity between reference and target \roser{samples} to 
obtain global and per-pair distances, based on CoverID, KL divergence, CLAP and DEfNet. It can estimate data replication with a proportion higher than $10\%$ (3 seconds), but in most of the examined scenarios, it is sensible to $5\%$ of replication. Per-pair distances are highly beneficial for detecting close pairs, outliers and suspicious cases with potential data replication. 
Considering that replication detection requirements may vary depending on the evaluation, users are left to set their replication threshold.
In addition, MiRA is model-independent as no information about the model architecture or its characteristics is necessary. The evaluation is conducted directly with the training (\textbf{reference}) and generated samples (\textbf{target}) of the analysed generative model.

%%%%%%%%%%%%%%%%%%%%%%
% Employment
%%%%%%%%%%%%%%%%%%%%%%
However, designating a baseline value is encouraged to accurately interpret the music similarity between the reference and target samples. We propose a third comparison group of samples (\textbf{control}) based on songs related to the reference songs but unseen by the model (e.g. shared music genre). Again, this is a decision for the users conditioned to their evaluation scope.  %\textcolor{red}{comment on genre} 
Note that using a control group allows us to understand and interpret the results obtained by acting as the baseline similarity level of independent songs with a shared characteristic. 

\begin{figure}[h]
 \centerline{
 \includegraphics[width=0.85\columnwidth]{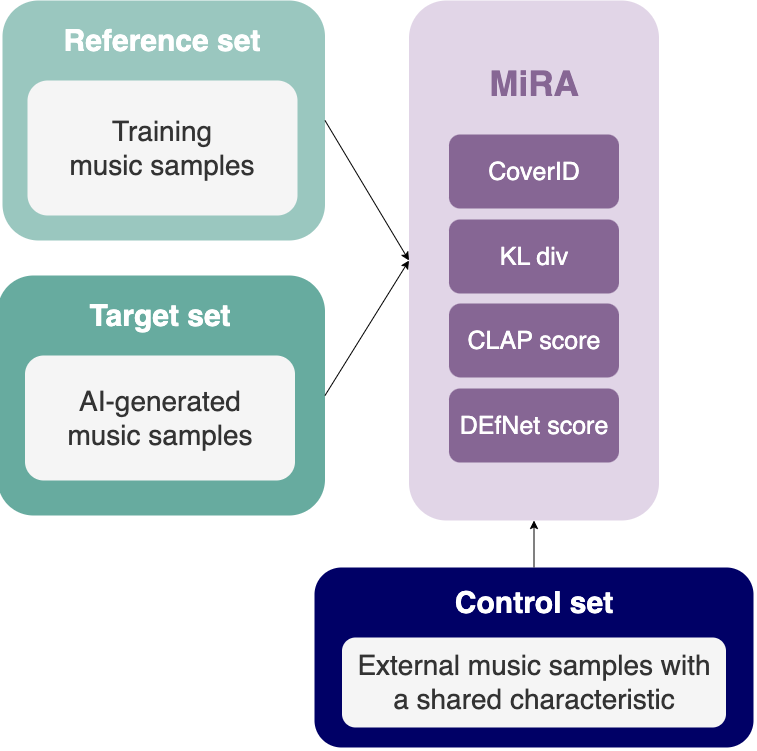}}
 \caption{\ourtool's~ structure scheme.}
 \label{fig:mira}
\end{figure}

The complete structure of the implemented system is depicted in Figure \ref{fig:mira}. 
We release MiRA 
as an open-source tool, built into a PyPI package\footnote{\url{https://pypi.org/project/mira-sim/}}. Together with the code, we provide examples 
and best practice recommendations for using this methodology. With the release of MiRA, we hope to enhance transparency in music generation models and data replication assessment.
\vspace{-0.15cm}

\section{Discussion and Conclusions}
\label{sec:conclusions}
This work focused on validating the use of \roser{music} similarity metrics for \roser{assessing data replication in AI-generated music. We hypothesise that similarity metrics are effective in estimating data replication. 
Therefore, we framed the scope of our study to exact data replication in music samples, while conducting a controlled forced-replication experiment with synthetic data.}

\roser{We examined five diverse audio-based metrics: four standard metrics (CoverID, KL divergence, CLAP and FAD) and a novel one (DEfNet). Our results indicate that four of the five studied metrics can detect data replication to a certain extent.}
Instead, FAD based on CLAP music embeddings \roser{presented} an opposite behaviour \roser{compared to the other} metrics. Higher similarity is obtained for the baseline group and FAD shows unstable trends throughout the diverse music genres. Thus, we do not find it suitable for our case study. However, it must be acknowledged that the recent publication by Gui et al. \cite{Gui2023} offered multiple classifiers to compute FAD. There is the possibility that we did not consider the appropriate classifier for our task. Thus, we should consider exploring other classifiers before determining the \roser{validity} of FAD \roser{in detecting replication} in music. 

Regarding the other four metrics, our results show interesting insights. First, we find CoverID to be sensible to different replication degrees, establishing a robust threshold level at $10\%$ of replication. \roser{Furthermore, in some of the studied cases, replication sensitivity is lowered to $5\%$ of replication.} This is a substantial finding to validate the suitability of metrics oriented to specific music characteristics, such as tempo, structure and composition. 

Next, we observe that KL divergence can \roser{be sensitive to} replication as pairwise comparison between baseline and degrees of replication is statistically significant. Nevertheless, the other pairwise results reveal that KL divergence is ineffective for differentiating between replication degrees. We consider this an unexpected turnout in our analysis. 

Considering CLAP and DEfNet scores, both embedding-based metrics, our experiment validates their suitability to detect data replication. Not only do they show robustness by increasing their similarity value parallel to the replication degrees \roser{(i.e. higher similarity for higher level of replication)}, but they also show high sensitivity for different degrees of replication. All results suggest their sensitivity might be higher than we envisioned and might be able to detect replication in smaller samples \roser{(i.e. < 1.5 seconds)}. 

As a result of these findings, we \roser{achieve} our second goal \roser{within} the scope of this research: to build an open model-agnostic tool based on music similarity metrics on raw audio. \roser{In this article, }we have introduced the MiRA tool, leveraging the four validated similarity metrics, which can be used to evaluate any music-generative model with audio output. \roser{MiRA does not require any information about the model architecture or its characteristics. Instead, similarity evaluation relies on comparing reference and target samples. }

By introducing the MiRA tool, we are contributing to the research gap of lack of evaluation methodologies directly assessing potential data replication in AI-generated music. Our study validates the use of similarity metrics to estimate training data replication. We intend to encourage the open evaluation of music generation models by researchers, developers and users concerning data replication. In addition, our research strives for the importance of ethical, social, legal and economic consequences of generative AI in the music domain, together with the need to address their risks and issues. 

\subsection{Limitations and Future Work}
Despite our contribution to advance towards data replication assessment with music similarity metrics, there are multiple opportunities to complement our investigation. 

First, we limited the scope of our experimental approach to assessing the use of \roser{different} music similarity metrics for exact data replication,  
\roser{consequently reducing the definition of plagiarism to exact replication of fragments in the training set. We followed such an approach to validate our hypothesis and ensure an attainable method to address this issue. While this reduced scope could potentially be solved using audio fingerprinting strategies \cite{Cano2005}, 
we believe that by employing a diverse range of metrics we can provide a more comprehensive assessment of data replication.}

\roser{Framing our aim to exact data replication also} introduced a limitation in considering typical perturbations that music samples experience when training the model or during the model procedure to generate a music sample. Thus, it would be a key point for future work to validate the robustness of these metrics towards typical data augmentation techniques, such as pitch shifting and reverberation. Proving them to be robust would also enhance the capabilities of MiRA for detecting potential replication in AI-generated music. At the same time, we intend to expand the abilities of MiRA for data replication by incorporating complementary metrics, if necessary.   

In addition, our experimental process was limited to the high computational costs of some of the metrics. In particular, we faced significantly large amounts of time to compute FAD and KL divergence. This is a relevant concern as we want MiRA to be an open tool that can be used by any researcher or user. Thus, considering the computational capacity required to compute the integrated metrics within is a relevant issue in our research. 

Another limitation is the type of data that we use. We base our experiment on synthetic data despite our goal being oriented to AI-generated music. We must use synthetic data with a controlled percentage of replication to guarantee and assess the capabilities of detection \roser{and sensitivity }of music similarity metrics. However, we would like to test the validity of the introduced tool when used in a generation context. To do so, we require not only a generative model but its details on training data and generation samples. \roser{We plan to expand our research in with AI-generated content in upcoming studies.}

\section{Acknowledgments}
The authors would like to thank Gaëtan Hadjeres and William Thong from Sony AI, as well as Dmitry Bogdanov and Pablo Alonso-Jiménez from the Music Technology Group at Universitat Pompeu Fabra, for their insightful discussions and valuable feedback throughout the development of this research.

% \textbf{Do not include in your submission, only in your camera ready version}. This section can be used to refer to any
% individuals or organizations that should be acknowledged in this paper. This section does \textit{not} count towards the 
% page limit for scientific content.

\section{Ethics Statement}
The late rapid popularity growth of generative AI in the music domain brings significant ethical implications. The main challenges are linked to the role of AI within music creative processes, such as composition, potential misappropriation of data in AI-generated music, inquiries on the novelty of generations, derived authorship attribution, effects on intellectual property rights and sustainability of current business models. In addition, there are notable concerns about the cultural bias in these systems and their environmental impact. 

Our research focused on the issue of assessing potential data replication in AI-generated music. We observed a lack of evaluation methodologies to examine replication in raw audio. We contributed to this issue by proposing a methodology based on audio-based music similarity metrics. We demonstrated its effectiveness and provided an open tool to evaluate AI-generated music. Our introduced approach is contributing to the transparency of music generation algorithms. 

Despite the positive contribution of our investigation, we must be critical of some methodological aspects of our work. Our principal ethical concern falls under the type of data used to conduct our forced-replication experiment. In particular, we employ an internal dataset created with Spotify previews (30-second samples of music). Even if these practices are common in the ISMIR community, we see the need for guidelines for the legal assessment of MIR data included in datasets, incorporating country dependencies, origin and intended use, personal data involved (from artists and listeners) and potential future consequences\footnote{We refer to a recently documented example of research vs legal clash linked to algorithmic auditing in the music domain  \url{https://www.rollingstone.com/pro/features/spotify-teardown-book-streaming-music-790174/}}.

% For bibtex users:
\bibliography{ISMIR.bib}

\begin{thebibliography}{10}
\providecommand{\url}[1]{#1}
\csname url@samestyle\endcsname
\providecommand{\newblock}{\relax}
\providecommand{\bibinfo}[2]{#2}
\providecommand{\BIBentrySTDinterwordspacing}{\spaceskip=0pt\relax}
\providecommand{\BIBentryALTinterwordstretchfactor}{4}
\providecommand{\BIBentryALTinterwordspacing}{\spaceskip=\fontdimen2\font plus
\BIBentryALTinterwordstretchfactor\fontdimen3\font minus \fontdimen4\font\relax}
\providecommand{\BIBforeignlanguage}[2]{{%
\expandafter\ifx\csname l@#1\endcsname\relax
\typeout{** WARNING: IEEEtran.bst: No hyphenation pattern has been}%
\typeout{** loaded for the language `#1'. Using the pattern for}%
\typeout{** the default language instead.}%
\else
\language=\csname l@#1\endcsname
\fi
#2}}
\providecommand{\BIBdecl}{\relax}
\BIBdecl

\bibitem{Carnovalini2020}
F.~Carnovalini and A.~Rodà, ``Computational creativity and music generation systems: An introduction to the state of the art,'' \emph{Frontiers in Artificial Intelligence}, vol.~3, 2020.

\bibitem{Gomez2018}
\BIBentryALTinterwordspacing
E.~Gómez, M.~Blaauw, J.~Bonada, P.~Chandna, and H.~Cuesta, ``Deep learning for singing processing: Achievements, challenges and impact on singers and listeners,'' \emph{ArXiv}, 2018. [Online]. Available: \url{https://arxiv.org/abs/1807.03046v1}
\BIBentrySTDinterwordspacing

\bibitem{Strum2019}
B.~L.~T. Sturm, M.~Iglesias, O.~Ben-Tal, M.~Miron, and E.~Gómez, ``Artificial intelligence and music: Open questions of copyright law and engineering praxis,'' \emph{Arts}, vol.~8, p. 115, 2019.

\bibitem{Yin2022}
Z.~Yin, F.~Reuben, S.~Stepney, and T.~Collins, ``Measuring when a music generation algorithm copies too much: The originality report, cardinality score, and symbolic fingerprinting by geometric hashing,'' \emph{SN Computer Science}, vol.~3, 2022.

\bibitem{Carlini2023a}
N.~Carlini, J.~Hayes, M.~Nasr, M.~Jagielski, V.~Sehwag, F.~Tramer, B.~Balle, D.~Ippolito, and E.~Wallace, ``Extracting training data from diffusion models,'' in \emph{32nd USENIX Security Symposium (USENIX Security 23)}, 2023, pp. 5253--5270.

\bibitem{Carlini2023b}
N.~Carlini, D.~Ippolito, M.~Jagielski, K.~Lee, F.~Tramer, and C.~Zhang, ``Quantifying memorization across neural language models,'' \emph{ArXiv}, 2023.

\bibitem{bralios2023}
D.~Bralios, G.~Wichern, F.~G. Germain, Z.~Pan, S.~Khurana, C.~Hori, and J.~L. Roux, ``Generation or replication: Auscultating audio latent diffusion models,'' \emph{ArXiv}, 2023.

\bibitem{Wang2023}
H.~Wang, ``Authorship of artificial intelligence-generated works and possible system improvement in china,'' \emph{Beijing Law Review}, vol.~14, pp. 901--912, 2023.

\bibitem{Somepalli2022}
G.~Somepalli, V.~Singla, M.~Goldblum, J.~Geiping, and T.~Goldstein, ``Diffusion art or digital forgery? investigating data replication in diffusion models,'' \emph{ArXiV}, 2022.

\bibitem{Barnett2023}
J.~Barnett, ``The ethical implications of generative audio models: A systematic literature review,'' \emph{AIES 2023 - Proceedings of the 2023 AAAI/ACM Conference on AI, Ethics, and Society}, pp. 146--161, 2023.

\bibitem{agostinelli2023}
A.~Agostinelli, T.~I. Denk, Z.~Borsos, J.~Engel, M.~Verzetti, A.~Caillon, Q.~Huang, A.~Jansen, A.~Roberts, M.~Tagliasacchi, M.~Sharifi, N.~Zeghidour, and C.~Frank, ``Musiclm: Generating music from text,'' \emph{ArXiv}, 2023.

\bibitem{Chen2023}
K.~Chen, Y.~Wu, H.~Liu, M.~Nezhurina, T.~Berg-Kirkpatrick, and S.~Dubnov, ``{MusicLDM}: Enhancing novelty in text-to-music generation using beat-synchronous mixup strategies,'' \emph{ArXiv}, 2023.

\bibitem{Xiong2023}
Z.~Xiong, W.~Wang, J.~Yu, Y.~Lin, and Z.~Wang, ``A comprehensive survey for evaluation methodologies of {AI}-generated music,'' \emph{ArXiv}, 2023.

\bibitem{batlle2023}
R.~Batlle-Roca, E.~G{\'o}mez, W.~Liao, X.~Serra, and Y.~Mitsufuji, ``Transparency in music-generative {AI}: A systematic literature review,'' \emph{Research Square preprint}, 2023.

\bibitem{Barnett2024}
J.~Barnett, H.~F. Garcia, and B.~Pardo, ``Exploring musical roots: Applying audio embeddings to empower influence attribution for a generative music model,'' \emph{arXiv}, 2024.

\bibitem{flores2023}
H.~F. {Flores Garcia}, P.~Seetharaman, R.~Kumar, and B.~Pardo, ``Vampnet: Music generation via masked acoustic token modeling,'' in \emph{Proceedings of the 24th International Society for Music Information Retrieval Conference, {ISMIR} 2023, Milan, Italy}, 2023.

\bibitem{Spijkervet2021}
J.~Spijkervet and J.~A. Burgoyne, ``Contrastive learning of musical representations,'' \emph{Proceedings of the 22nd International Society for Music Information Retrieval Conference, {ISMIR} 2021, Online}, 2021.

\bibitem{wu2023}
Y.~Wu, K.~Chen, T.~Zhang, Y.~Hui, T.~Berg-Kirkpatrick, and S.~Dubnov, ``Large-scale contrastive language-audio pretraining with feature fusion and keyword-to-caption augmentation,'' \emph{ArXiv}, 2023.

\bibitem{Serra2009}
J.~Serrà, X.~Serra, and R.~Andrzejak, ``Cross recurrence quantification for cover song identification,'' \emph{New Journal of Physics}, vol.~11, 2009.

\bibitem{Serra2008}
J.~Serrà, E.~Gómez, P.~Herrera, and X.~Serra, ``Chroma binary similarity and local alignment applied to cover song identification,'' \emph{IEEE Transactions on Audio, Speech, and Language Processing}, vol.~16, no.~6, pp. 1138--1151, 2008.

\bibitem{Serr2008TransposingCR}
J.~Serr{\`a}, E.~G{\'o}mez, and P.~Herrera, ``Transposing chroma representations to a common key,'' \emph{IEEE CS Conference on The Use of Symbols to Represent Music and Multimedia Objects}, 2008.

\bibitem{Hoffman2008}
M.~Hoffman, D.~Blei, and P.~Cook, ``Content-based musical similarity computation using the hierarchical dirichlet process.'' in \emph{Proceedings of the 9th International Society for Music Information Retrieval Conference, {ISMIR} 2008, Philadelphia, USA}, 2008.

\bibitem{Schnitzer2010}
D.~Schnitzer, A.~Flexer, G.~Widmer, and M.~Gasser, ``Islands of gaussians: The self organizing map and gaussian music similarity features,'' in \emph{Proceedings of the 11th International Society for Music Information Retrieval Conference, {ISMIR} 2010, Utrecht, Netherlands}, 2010.

\bibitem{evans2024}
Z.~Evans, C.~Carr, J.~Taylor, S.~H. Hawley, and J.~Pons, ``Fast timing-conditioned latent audio diffusion,'' \emph{ArXiv}, 2024.

\bibitem{Koutini2022}
K.~Koutini, J.~Schl{\"{u}}ter, H.~Eghbal{-}zadeh, and G.~Widmer, ``Efficient training of audio transformers with patchout,'' in \emph{Interspeech 2022, 23rd Annual Conference of the International Speech Communication Association, Incheon, Korea}.\hskip 1em plus 0.5em minus 0.4em\relax {ISCA}, 2022, pp. 2753--2757.

\bibitem{kreuk2023}
F.~Kreuk, G.~Synnaeve, A.~Polyak, U.~Singer, A.~Défossez, J.~Copet, D.~Parikh, Y.~Taigman, and Y.~Adi, ``Audiogen: Textually guided audio generation,'' 2023.

\bibitem{copet2023}
J.~Copet, F.~Kreuk, I.~Gat, T.~Remez, D.~Kant, G.~Synnaeve, Y.~Adi, and A.~Défossez, ``Simple and controllable music generation,'' \emph{ArXiv}, 2023.

\bibitem{alonso2020}
P.~Alonso-Jim{\'e}nez, D.~Bogdanov, J.~Pons, and X.~Serra, ``Tensorflow audio models in {Essentia},'' in \emph{International Conference on Acoustics, Speech and Signal Processing ({ICASSP})}, 2020.

\bibitem{Kilgour2019}
K.~Kilgour, M.~Zuluaga, D.~Roblek, and M.~Sharifi, ``{Fréchet Audio Distance: A Reference-Free Metric for Evaluating Music Enhancement Algorithms},'' in \emph{Proc. Interspeech 2019}, 2019, pp. 2350--2354.

\bibitem{Gui2023}
A.~Gui, H.~Gamper, S.~Braun, and D.~Emmanouilidou, ``Adapting frechet audio distance for generative music evaluation,'' \emph{ArXiv}, 2023.

\bibitem{hershey2017}
S.~Hershey, S.~Chaudhuri, D.~P.~W. Ellis, J.~F. Gemmeke, A.~Jansen, R.~C. Moore, M.~Plakal, D.~Platt, R.~A. Saurous, B.~Seybold, M.~Slaney, R.~J. Weiss, and K.~Wilson, ``Cnn architectures for large-scale audio classification,'' \emph{ArXiv}, 2017.

\bibitem{KruskalWallis1952}
W.~H. Kruskal and W.~A. Wallis, ``Use of ranks in one-criterion variance analysis,'' \emph{Journal of the American Statistical Association}, vol.~47, no. 260, pp. 583--621, 1952.

\bibitem{Cano2005}
P.~Cano and E.~Batlle, ``A review of audio fingerprinting,'' \emph{Journal of VLSI Signal Processing}, vol.~41, pp. 271--284, 11 2005.

\end{thebibliography}

\end{document}